\documentclass[preprint]{revtex4}
\usepackage[latin1]{inputenc}
\usepackage{graphicx,epstopdf}
\usepackage{subfigure}
\usepackage{epsfig}
\usepackage{bigints}
\usepackage[usenames]{color}
\usepackage{slashed}

\newcommand{\beqa}{\begin{eqnarray}}
\newcommand{\eeqa}{\end{eqnarray}}
\newcommand{\beq}{\begin{equation}}
\newcommand{\eeq}{\end{equation}}

\newcommand{\abs}[1]{\lvert#1\rvert}

\begin{document}

\title{The QCD Kondo phase in quark stars}

\author{R.\ Fariello{$^{\dag\,\ddag}$}, J.C.  Mac\'{\i}as{$^\ddag$} and F.\ S.\ Navarra{$^\ddag$}}

\address{{$^\dag$}Departamento de Ci\^encias da Computa\c{c}\~ao, Universidade Estadual de Montes Claros,
	Avenida Rui Braga, sn, Vila Mauriceia, 39401-089, Montes Claros, Minas Gerais, Brazil}
\address{{$^\ddag$}Instituto de F\'isica, Universidade de S\~ao Paulo,
	Rua do Mat\~ao Travessa R, 187, 05508-090 S\~ao Paulo, SP, Brazil}

\begin{abstract}
We study light ($u$, $d$)\ quark matter with charm impurities. These impurities are added to the Lagrangian density. 
We derive the equation of state (EOS) of this kind of quark matter, which 
contains a Kondo phase. We explore this EOS and study the structure of stars, identifying the effects of the Kondo phase.
Solving the  TOV equations and computing the mass-radius diagram, we find that the presence of a Kondo phase   leads to 
 smaller and lighter stars.

\end{abstract}

\maketitle

\section{Introduction}

A long standing question in the theory of compact stars \cite{glend,hebel,emmi,kojo,fukojo,drago,fkv16,kfbv,fkv14} is:  
Are there quark stars? This question has been around for decades and it has received a
renewed attention after the appearance of new measurements of masses of astrophysical compact objects \cite{demorest,anton,vanker}.
These measurements suggest that stellar objetcs may have large masses, such as $(1.97 \pm 0.04)\, M_{\odot}$ \cite{demorest},  
$(2.01 \pm 0.04)\, M_{\odot}$ \cite{anton} or even $(2.4 \pm 0.12)\, M_{\odot}$ \cite{vanker}. In principle larger masses imply 
larger baryon densities in the core of the stars and we expect very dense hadronic matter to be in a quark gluon plasma (QGP)\ phase. 
On the other hand, most of the equations of state based on quark degrees of freedom are too soft to support heavy stars.

The existence of quark stars depends ultimately on the details of the equation of state of cold quark matter. 
According to most models, deconfined quark matter should
be formed at baryon densities in the range $\rho_B = 2 \rho_0\textup{--}5 \rho_0$, where $\rho_0$ is the ordinary nuclear matter 
baryon density. Since at low temperatures and high baryon densities we can not rely on lattice QCD calculations, the quark matter 
equations of state must be derived from models. Many of them are based on the MIT bag model \cite{mit} or on the 
Nambu-Jona-Lasinio (NJL)\ model \cite{nambu}. At very high baryon densities there are constraints derived from perturbative 
QCD calculations \cite{fkv16,kfbv,fkv14,pqcd}.

The description of cold quark matter is not unique and it may (or may not)\ contain specific QCD features such as color 
superconductivity, diquarks, or a Gribov-Zwanziger phase \cite{grizwan}. One of these QCD features is the QCD Kondo phase.  
Recently \cite{yasui16} the Kondo ef\/fect has been studied in the context of quark matter. In \cite{yasui16} it was pointed out 
that the Kondo ef\/fect occurs when a system has i)\ heavy impurities, ii)\ a Fermi surface of fermions, iii)\ quantum f\/luctuations, and 
iv)\ non-Abelian interactions. All these features are present in a dense and cold light quark system with some heavy quarks as impurities. 
This kind of quark matter was called Kondo phase in \cite{yasui16} and its existence in the core of dense stars may change 
the thermodynamic and transport properties of the stellar medium. 

In compact stars the heavy impurities are charm quarks in low concentration. Charm can be produced in 
strange quark stars by neutrino interactions. A constant neutrino f\/lux traverses the star. 
During their interactions with quark matter, neutrinos emit a $W^+$ ($\nu_e \rightarrow W^+ \, e^-$), 
which is absorbed by an $s$ quark (or to a lesser extent by a $d$ quark), which turns into a charm quark 
($W^+ \, s \rightarrow c$). After being produced the $c$ quark can decay back to an $s$ quark but Pauli 
blocking will reduce the ef\/f\/iciency of this reaction.

The existence of charm in quark stars was first investigated in Ref. \cite{weber}. At the time the conclusion 
was  that this kind of star would be unstable. Very recently \cite{newfraga} this question was addressed again in the context of 
perturbative QCD (pQCD). The authors investigated the effects of charm quarks in the equation of state  
for large values of the quark chemical potential, where pQCD should be reliable. The radial stability analysis suggested that 
this star would be unstable.  Even though charm stars probably do not exist, it is conceivable that some 
f\/inite amount of charm will always be present in the star and this may be enough to generate the Kondo phase.

In early works the QCD Kondo ef\/fect was studied with the perturbative renormalization group equation obtained at
the one-loop level \cite{yasui15}. In \cite{yasui16} the ground state of the quark matter with heavy impurities was investigated with a  
non-perturbative mean f\/ield approach. The authors used the following Lagrangian density \cite{yasui16}
\beq 
\mathcal{L} =  \Bar{\psi} i \slashed{\partial}\psi 
            + \mu\, \Bar{\psi} \gamma^0 \psi 
						+  \Bar{\Psi} i \slashed{\partial}\Psi  
						-  m_Q  \Bar{\Psi} \Psi
            - G_c\, (\Bar{\psi} \gamma^{\mu} T^a \psi)(\Bar{\Psi} \gamma_{\mu} T^a \Psi)
\label{lagr}
\eeq
where $\psi$ and $\Psi$ represent the light and heavy quark f\/ields respectively. In the heavy quark limit 
the latter can be replaced by $\Psi \rightarrow \Psi_v = \frac{1}{2} (1 + \slashed{v})\, e^{i m_Q v\cdot x} \Psi$, where $v$ is the 
velocity of the heavy quark. The coupling strength $ G_c $ in the interaction part of $ \mathcal{L} $ is positive and has dimensions 
of inverse mass square. The interaction term has the color structure $ \lambda^a\lambda^a $ as prescribed by the one-gluon exchange. 
The values of the coupling and the cutoff parameter which we have denoted as $ \Lambda $ are taken from the usual NJL model 
(for $N_f = 2$):\ $G_c\,\Lambda^2 = (9/2)2.0$ and $\Lambda = 0.65$ GeV. These numbers are chosen so as to describe the pion decay 
constant and the quark 
condensate density. We have suppressed the flavor index for the f\/ields $\psi$ and we have assumed that they have the same chemical 
potential $\mu$. In \cite{yasui16} it was assumed that the heavy quarks are spatially uniformly distributed within the light quark 
matter and the density of heavy quarks is large so that the averaged distance between heavy quarks is small when compared to a typical 
coherence length of the QCD Kondo ef\/fect. The above Lagrangian is treated in the mean-f\/ield approach and the four-quark term 
appearing in (\ref{lagr}) can be factorized, giving rise to condensates such as, for example, $\langle\Bar{\psi} \Psi_v\rangle$. 
In momentum space the (bilinear)\ mean-f\/ield Lagrangian appears as follows \cite{yasui16}
\beqa
\mathcal{L}^{MF} &=& \Bar{\psi}\slashed{k}\psi 
                  + \mu\, \Bar{\psi} \gamma^0 \psi 
									+  \Bar{\Psi}_v v\cdot k \Psi_v  
                  - \lambda\,( {\Psi}^{\dagger}_v  \Psi_v - n_Q)
                  + \Delta \Bar{\Psi}_v \frac{1+\gamma^0}{2} (1 + \Hat{k}\cdot\Vec{\gamma})\psi \nonumber \\
                 &+& \Delta^* \Bar{\psi} (1 + \Hat{k}\cdot\Vec{\gamma}) \frac{1+\gamma^0}{2} \Psi_v 
                  - \frac{8\, N_f}{G_c}\,\abs{\Delta }^2 
\label{lagrmf}
\eeqa
where the term weighted by means of the Lagrange multiplier $\lambda$ was added to
impose the constraint of number 
conservation of the heavy quarks and $n_Q$ is the averaged heavy quark density. This
constraint is required since the 
heavy quark total number must be f\/ixed, on average.  As it was emphasized in Ref.     
\cite{yasui17}, since $(- \lambda)$ is the coefficient of $\Psi^{\dagger}_v \Psi_v$ one
might formally 
interpret it as the chemical potential of heavy quarks.  
Then, it is the  chemical potential  for the redefined heavy-quark field $\Psi_v$ rather
than the original heavy-quark field $\Psi$. In other words, a nonzero $\lambda$ can be
regarded as {\it the energy necessary to put a virtual component  of a heavy quark into the system}.
Even though $\lambda$ is not the real heavy quark chemical potential, it is useful to treat it
as if it were. As it will be seen, there is a strong correlation between $(- \lambda) $ and the
number of heavy quarks $n_Q$.  For a fixed light quark density (fixed $\mu$), $n_Q$ increases when
$(-\lambda)$ increases, as we would expect for a chemical potential.

The quantity $\Delta$ is a complex number associated with the gap function, 
which is def\/ined as \cite{yasui16}:
\beq
\Delta_{\delta \alpha} = \frac{G_c}{2} \langle\Bar{\psi}_{\alpha} \Psi_{v \delta}\rangle
= \Delta\, ( \tfrac{1+\gamma^0}{2}(1 - \Hat{k}\cdot\Vec{\gamma}) )_{\delta \alpha}
\label{gap}
\eeq

\section{The equation of state and the Kondo phase}

From  the Lagrangian (\ref{lagrmf})\ we can derive the thermodynamic potential, which is given by \cite{yasui16}:
\beq 
\Omega (T,\mu ,\lambda ) = \frac{N_c}{\pi ^2} \int_0^{\Lambda } k^2\, f(T,\mu ,\lambda ,k) \, dk 
	+ \frac{8\, N_f}{G_c}\,\abs{\Delta }^2 
	- \lambda\, n_Q(T,\mu ,\lambda )
\label{Lanpot}	
\eeq
with $f$ being
\begin{multline}
f(T,\mu ,\lambda ,k) = 																				
-T\, \log{ \bigl( \exp{ \bigl( -\tfrac{E_+(\mu,\lambda ,k)}{T} \bigr) } + 1 \bigr) } \\
-T\, \log{ \bigl( \exp{ \bigl( -\tfrac{E_-(\mu,\lambda ,k)}{T} \bigr) } + 1 \bigr) } 
-T\, \log{ \bigl( \exp{ \bigl( -\tfrac{E(\mu,k)}{T} \bigr) } + 1 \bigr) }
\end{multline}
$E_+(\mu,\lambda ,k)$, $E_-(\mu,\lambda ,k)$ and $E(\mu,k)$ are the real parts of the Bogoliubov eigenenergies, which are
\beq
E_\pm(\mu ,\lambda ,k) = \tfrac{1}{2} \bigl( \pm \sqrt{\smash[b]{(k-\lambda -\mu )^2 
                       + 8\, N_f\, \abs{\Delta }^2 }} + k + \lambda -\mu \bigr),\quad
E(\mu,k) = k - \mu											
\eeq
Minimizing the potential with respect to the Lagrange multiplier $\lambda$, i.e., taking  
\beq
\frac{\partial\, \Omega (T,\mu ,\lambda )}{\partial \lambda}=0
\label{nheavy}
\eeq
we obtain an expression for the number density of heavy quarks:
\beq
n_Q(T,\mu ,\lambda ) = \frac{N_c}{\pi ^2} \int_0^{\Lambda } k^2\, \frac{\partial f(T,\mu ,\lambda ,k)}{\partial \lambda} \, dk  
\label{nqezao}
\eeq 
In the zero-temperature limit, \eqref{Lanpot} reduces to the following form:
\beq  
\Omega (\mu ,\lambda )
 = - \frac{\lambda \, N_c}{\pi ^2} \int_0^{\Lambda } k^2\, \frac{\partial f_0(\mu ,\lambda ,k)}{\partial \lambda} \, dk 
   + \frac{N_c}{\pi ^2} \int_0^{\Lambda } k^2\, f_0(\mu ,\lambda ,k) \, dk 
	 + \frac{8\, N_f}{G_c}\,\abs{\Delta }^2
\eeq
with
\beq
f_0(\mu ,\lambda ,k) = \theta (-k-\lambda +\mu -\sigma )\, E_+(\mu ,\lambda ,k) 
+\theta (-k-\lambda +\mu +\sigma )\, E_-(\mu ,\lambda ,k) 
+\theta (\mu -k)\, E(\mu,k)
\eeq
where $\theta$ is the unit step function, and where we def\/ine 
$\sigma = \sqrt{\smash[b]{(k-\lambda -\mu )^2 + 8\, N_f\, \abs{\Delta }^2 }}$. 
In the calculations, the parameters were taken from Ref.~\cite{yasui16}. The pressure
and energy density are given by
\beq
P(\mu ,\lambda ) = -\Omega (\mu ,\lambda )
\label{press}
\eeq
\beq
\varepsilon (\mu ,\lambda ) = -P(\mu ,\lambda ) + \mu \, n_q(\mu ,\lambda )
- \lambda \,  n_Q(\mu ,\lambda )
\label{eps}
\eeq
where $n_q$ is the number density of light quarks:
\beq
n_q(\mu ,\lambda )=-\frac{\partial\, \Omega (\mu ,\lambda )}{\partial \mu}
   =\frac{N_c \left(\lambda 
   \int_0^{\Lambda } k^2\,
   \frac{\partial ^{2} f_0(\mu ,\lambda ,k)}{\partial \mu\, \partial \lambda}
   \, dk-\int_0^{\Lambda } k^2\,
   \frac{\partial f_0(\mu ,\lambda ,k)}{\partial \mu} 
   \, dk\right)}{\pi ^2}
\label{nqzin}
\eeq

\begin{figure}[ht!]
\begin{center}
\subfigure[ ]{\label{fig:first1}
\includegraphics[width=0.488\textwidth]{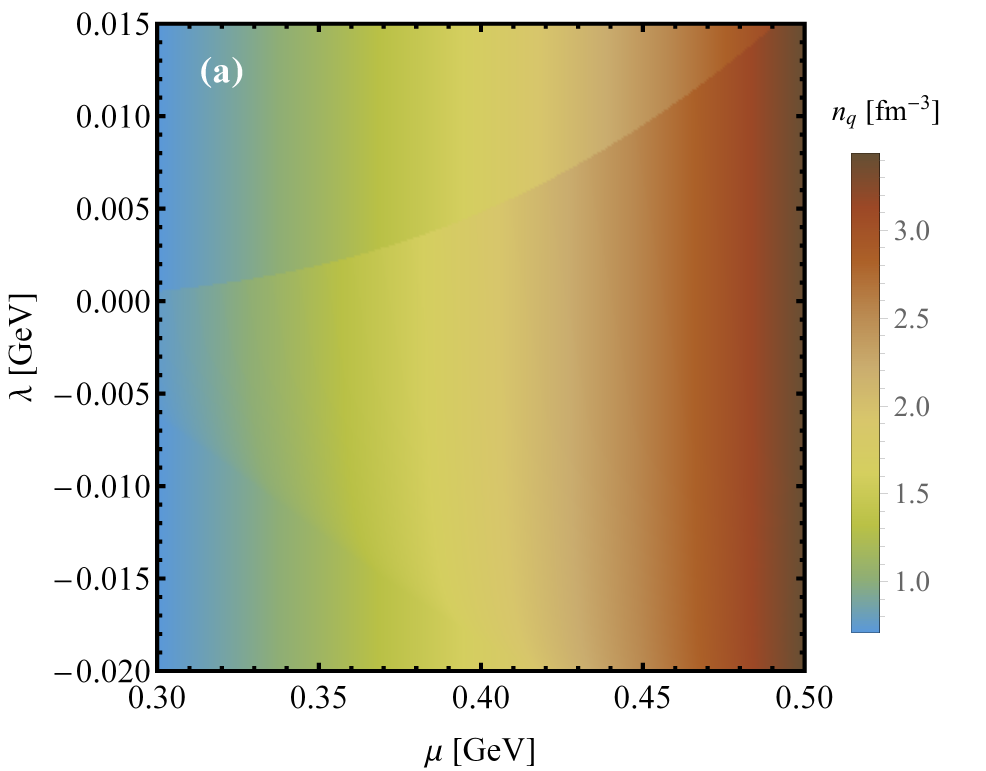}}\\
\subfigure[ ]{\label{fig:second1}
\includegraphics[width=0.488\textwidth]{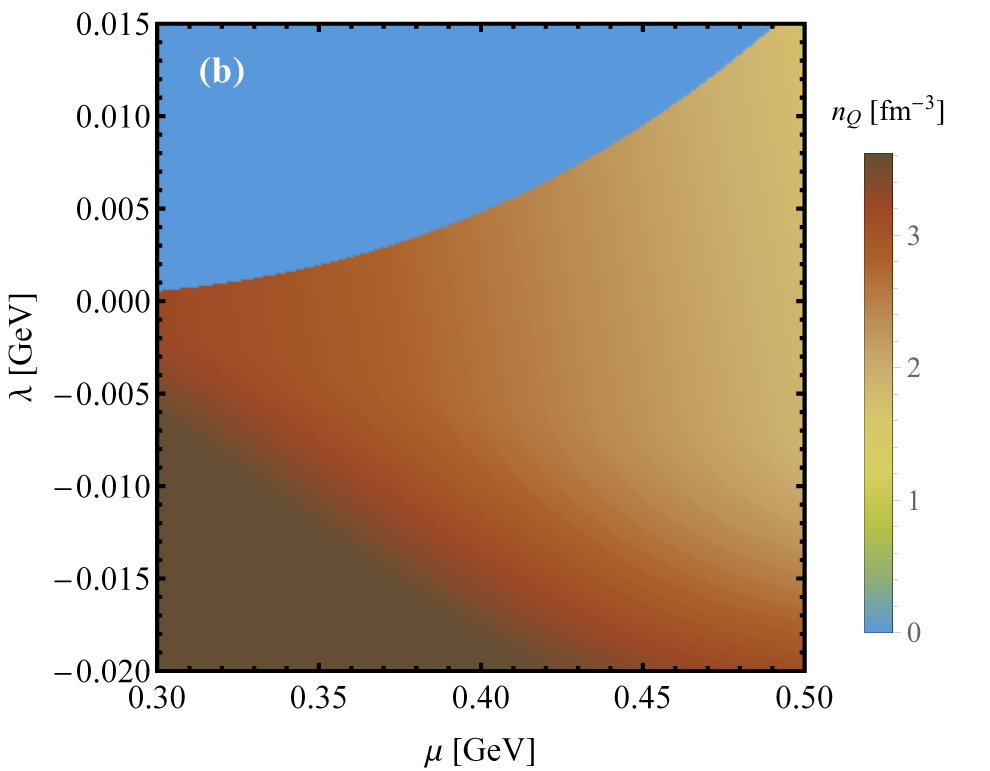}}\\
\subfigure[ ]{\label{fig:second1}
\includegraphics[width=0.488\textwidth]{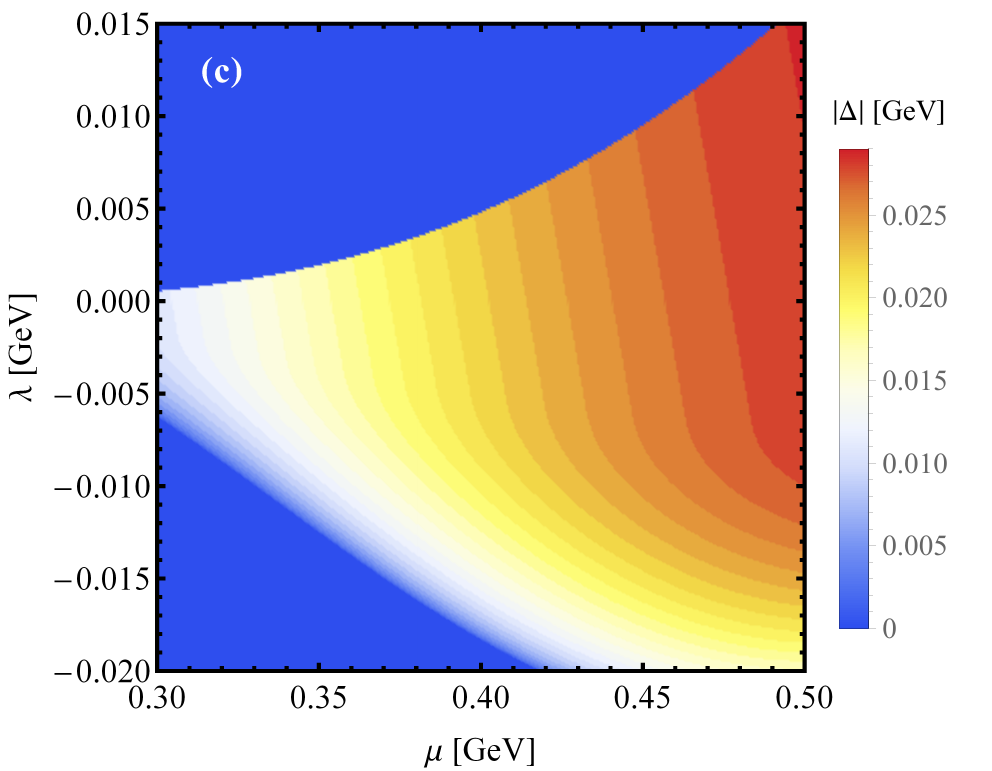}}
\end{center}
\caption{(a) Light quark density as a function of $\mu$ and $\lambda$; (b) the same as
(a) for the heavy quark density; (c) the same as (a) for the gap. }
\label{fig1}
\end{figure}

In Fig.~\ref{fig1} we show the quark densities $n_q$ and $n_Q$ and the gap $|\Delta|$ as a
function of $\mu$ and $\lambda$.  From the top picture (Fig. \ref{fig1}a) we see that the
light quark density is independent of $\lambda$. In contrast, the number of heavy quarks is strongly
sensitive to the value of $\mu$. As can be seen from Fig. \ref{fig1}b,  higher light quark
densities imply lower heavy quark densities. According to a naive expectation, the 
Kondo phase is the consequence of a non-vanishing gap $|\Delta|$, which in turn is a consequence
of non-vanishing impurities $n_Q$. Indeed, there is a correlation between $n_Q$ and
$|\Delta|$
seen in the top of Figs \ref{fig1}b and \ref{fig1}c:  $|\Delta|=0$  when $n_Q=0$. However, at
lower light quark densities, $|\Delta|$ is zero when $n_Q$ is maximal, as can be seen in bottom
left corner of Figs.~\ref{fig1}b and	\ref{fig1}c. 

\begin{figure}[t]
\epsfig{file=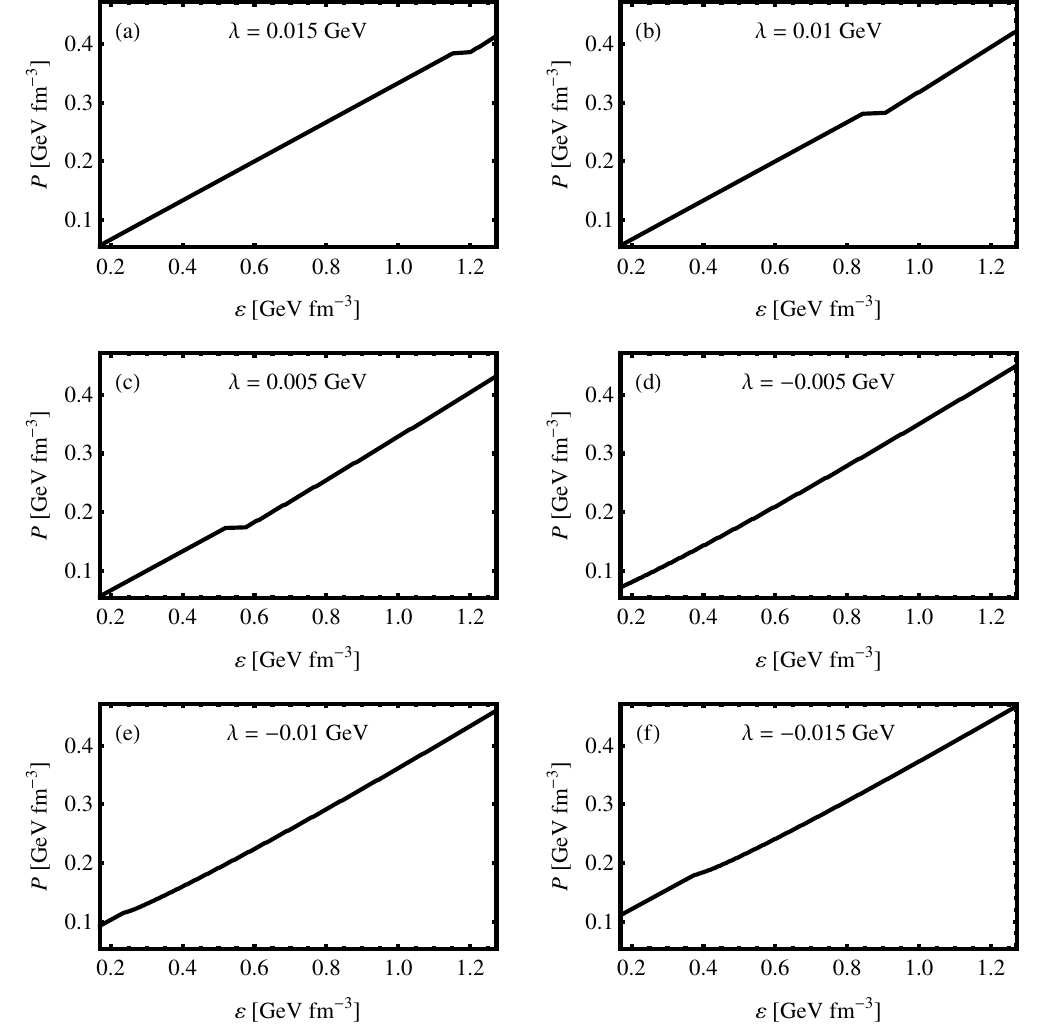,width=140mm}
\caption{Equation of state  obtained from (\ref{nqezao}), (\ref{press}), 
(\ref{eps}) and (\ref{nqzin}). }
\label{eos}
\end{figure}

The  equation of state is shown in Fig.~\ref{eos}, where we can see the pressure as a function of the   
energy density.  There is a connection between Figs. \ref{fig1} and \ref{eos}. For example, looking at 
the line of constant $\lambda = 0.005$ GeV both in Fig. \ref{fig1}b and  \ref{fig1}c we see that, at  
a certain critical value of $\mu$ both $n_Q$ and $|\Delta|$ suddenly start to grow.  This change of 
behavior corresponds to the appearance of  the plateau in Fig. \ref{eos}c, which marks the onset of the 
Kondo phase. 

\begin{figure}[t]
\epsfig{file=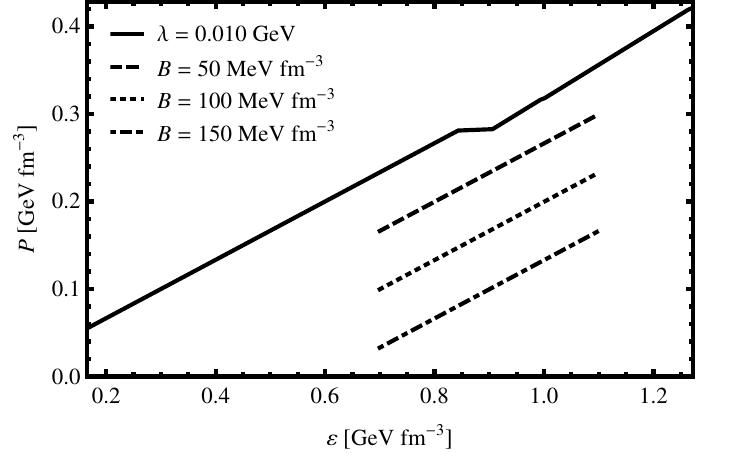,width=140mm}
\caption{Comparison between the EOS developed in this work with the MIT EOS.}
\label{mit}
\end{figure}

For
comparison, we also show the equation of state of the MIT bag model. This EOS 
for a QGP with $u$, $d$ and $s$ quarks (of equal masses $m_q$) has  pressure given by
\begin{equation}
  {P_{MIT}}=\sum_{q=u}^{d,s}\,{\frac{\gamma_{q}}{6\pi^{2}}}\int_{0}^{k_{F}} dk \,
  {\frac{{k}^{4}}{\sqrt{\smash[b]{m_{q}^{2}+k^{2}}}}}\, - B
  \label{pressmitz}
\end{equation}
and  energy density given by
\begin{equation}
  {\varepsilon_{MIT}}=\sum_{q=u}^{d,s}\,
  {\frac{\gamma_{q}}{2\pi^{2}}} \int_{0}^{k_{F}} dk \, k^{2}\, \sqrt{\smash[b]{m_{q}^{2}+k^{2}}}\,
  + B
  \label{energmitz}
\end{equation}
where $B$ is the bag constant, $m_q =10$ MeV and $\gamma_q=6$
is the statistical factor for quarks. The quark density is given by
\begin{equation}
  \rho_{MIT}=\sum_{q=u}^{d,s}\,{\frac{\gamma_{q}}{2\pi^{2}}}
  \int_{0}^{k_{F}} dk \, k^{2}
  \label{quarkdensmitz}
\end{equation}
which gives the highest occupied level $k_{F}$. Using  the baryon density 
$\rho_{B}={\tfrac{\rho_{MIT}}{3}}$, we get a simple expression for the Fermi momentum
\begin{equation}
  k_F = \pi^{2/3} \rho_B^{1/3}
  \label{barkf}
\end{equation}
which allows the calculation of the pressure and energy density as functions of the baryon density.
In Fig.  \ref{mit} we compare the MIT EOS with our equation of state for $\lambda = 0.01$ GeV. We can 
reach higher pressures with our model. We emphasize that, as can be seen from Fig.~\ref{eos}, all
the relevant values of $\lambda$ will generate EOS curves which will lie above the MIT curves.

\begin{figure}[t]
\epsfig{file=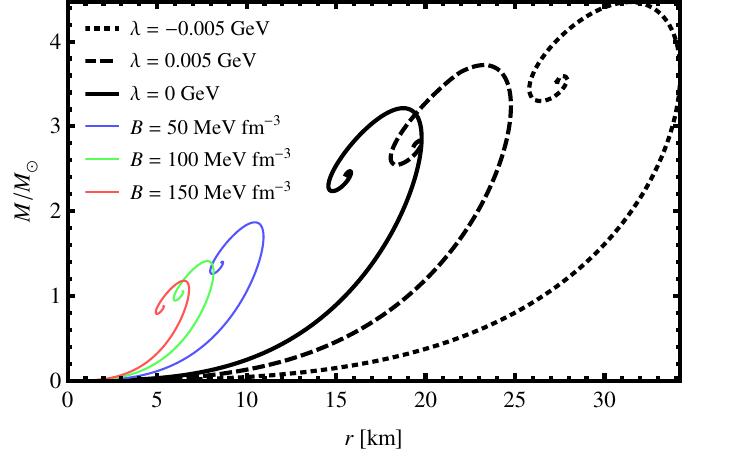,width=140mm}
\caption{Solution of the TOV equations (\ref{tov})}
\label{tov1}
\end{figure}

\begin{figure}[t]
\epsfig{file=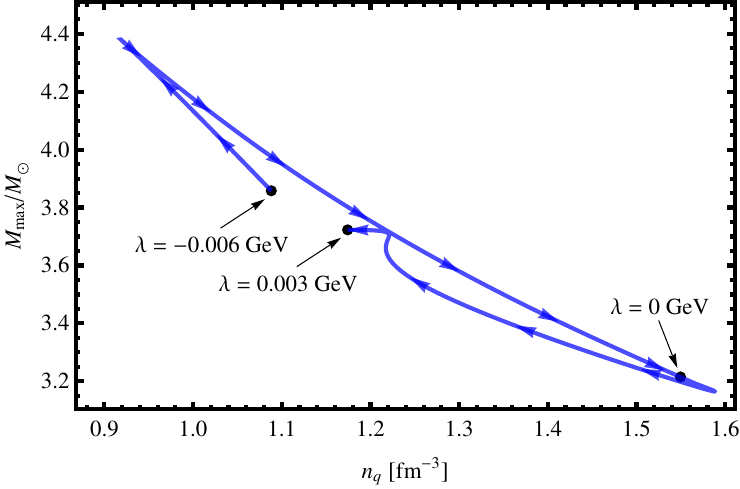,width=140mm}
\caption{Maximal star mass as a function of the light quark density $n_q$.}
\label{nq}
\end{figure}

\begin{figure}[t]
\epsfig{file=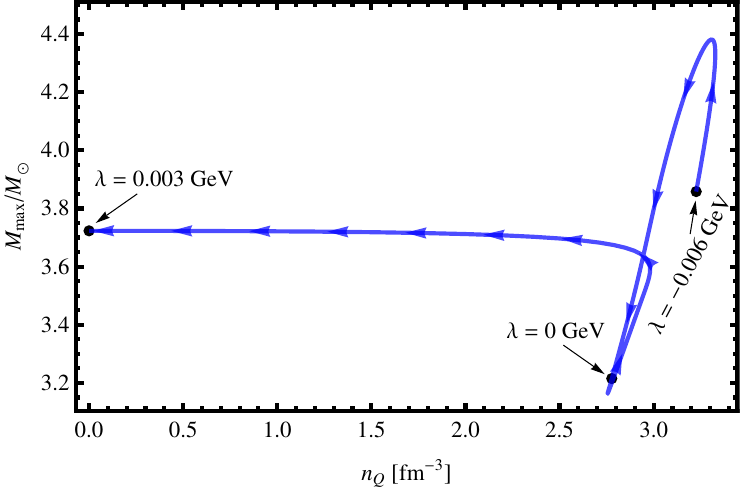,width=140mm}
\caption{Maximal star mass as a function of the heavy quark density $n_Q$.}
\label{nQ}
\end{figure}

\section{Stellar structure} 

As usual, in order to describe the structure of a static (non-rotating)\ compact star, the Einstein's f\/ield equations are 
solved for a medium with an isotropic relativistic f\/luid and in the case of a spherically symmetric metric tensor. 
Under these conditions, the Einstein equations imply the TOV system which becomes an integro-dif\/ferential equation 
for the pressure, $P$, as a function of the radius, $r$. The system reads:
\beq\label{tov}
P'(r)=-\frac{G\, M(r)\, \varepsilon (r)
   \left(\frac{P(r)}{\varepsilon
   (r)}+1\right) \left(\frac{4\, \pi\, 
   r^3 P(r)}{c^2
   M(r)}+1\right)}{c^2 r^2
   \left(1-\frac{2\, G\, M(r)}{c^2
   r}\right)},\quad
M(r)=\frac{4\, \pi  }{c^2}\int_0^r s^2
   \varepsilon (s) \, ds+M(0)
\eeq
where $G$ is Newton's gravitational constant and $c$ is the speed of light. Choosing the dimensionless variables 
$\tilde{P}(r)=\frac{P(r)}{P_0}$, $\tilde{\varepsilon }(r)=\frac{\varepsilon(r)}{\varepsilon _0}$, and 
$\tilde{M}(r)=\frac{M(r)}{M_\odot}$, following \cite{amj}, we have computed, for several f\/ixed pairs of input 
values $\left( P_0,\varepsilon _0\right )$, the total stellar mass (in solar masses) and then the corresponding 
stellar radius (in $\text{km}$) for all the relevant values of the parameter $\lambda$. Natural units have been adopted 
for all calculations.

In Fig.~\ref{tov1} we present some solutions of the TOV system of equations in the mass-radius diagram for both EOS models. 
In colored thin lines we show the results obtained with the
MIT bag model equation of state. In thick black lines those obtained with the model described here.
From Fig.~\ref{mit} we see that the model studied here generates harder equations of state
than the MIT. As a consequence it also generates heavier stars, as  shown by the thick black lines in Fig.~\ref{tov1}. 
The solid line shows the curve for $\lambda = 0$ GeV  which yields the
smallest value for $M_{max}$. All the other values of $\lambda$, both positive and negative, 
lead  right-lying curves in Fig.~\ref{tov1}. The origin of this non-trivial behavior is in
the lower panel of Fig.~\ref{fig1}. Indeed, from Fig.~\ref{fig1}c we see that the gap $\abs{\Delta}$
goes to zero when $\lambda$ moves away from zero both to positive and negative values. Since
non-zero values of $\abs{\Delta}$ are the signature of the Kondo phase, we can conclude that the
existence of a Kondo phase softens the equation of state and leads to lighter and smaller
stars.  In Fig.~\ref{nq} we show the dependence of $M_{max}$ on the  light  quark density, $n_q$.
In line with the results shown in Fig.~\ref{tov1}, we see that the smalest values of  $M_{max}$
occur for the $\lambda = 0 $, being larger for all other values of $\lambda$. 
In Fig.~\ref{nQ} we show the dependence	of $M_{max}$ on the  heavy quark density, $n_Q$. Here again,
the smaller maximal masses occur for $\lambda = 0$. Since heavy quarks are impurities in the present
model of quark matter, we expect to have $n_Q << n_q$.  
This condition will be satisfied for  small
values of $n_Q$, where, according to Fig.~\ref{nQ}, the dependence of $M_{max}$ on $n_Q$ is weak
and $M_{max} \to 3.7 M_{\odot}$.

\section{Conclusions}

We have evaluated the equation of state derived from the model developed in Ref.~\cite{yasui16}.  
We have applied this equation of state, which contains heavy quark impurities and has a Kondo phase,
to the study of quark stars. 
Solving the  TOV equations and computing the mass-radius diagram, we find that the existence of the
Kondo phase (when the gap $\abs{\Delta}$ is larger than zero) leads to softer equations of state and hence
to lighter and smaller stars.

\begin{acknowledgments}

This work was partially supported by the Brazilian funding agencies CAPES, CNPq and FAPESP. We thank Prof.\ Shigehiro Yasui 
for instructive discussions.

\end{acknowledgments}

\end{document}